\documentclass[conference]{IEEEtran}
\IEEEoverridecommandlockouts
\usepackage{cite}
\usepackage{amsmath,amssymb,amsfonts}
\usepackage{algorithmic}
\usepackage{graphicx}
\usepackage{xcolor}

\usepackage{hyperref}
\usepackage{newtxtext,newtxmath}

\usepackage[utf8]{inputenc}
\usepackage[english]{babel}

\begin{document}
\title{Multimedia Technology Applications and Algorithms: A Survey}
\author{\IEEEauthorblockN{Palak Tiwary}
\IEEEauthorblockA{Department of Computing Science\\
University of Alberta\\
Edmonton, Canada\\
Email: ptiwary@ualberta.ca}

\and
\IEEEauthorblockN{Sanjida Ahmed}
\IEEEauthorblockA{Department of Computing Science\\
University of Alberta\\
Edmonton, Canada\\
Email: sanjida1@ualberta.ca}}

\maketitle

\begin{abstract}
Multimedia related research and development has evolved rapidly in the last few years with advancements in hardware, software and network infrastructures. As a result, multimedia has been integrated into domains like Healthcare and Medicine, Human facial feature extraction and tracking, pose recognition, disparity estimation, etc. This survey gives an overview of the various multimedia technologies and algorithms developed in the domains mentioned.
\end{abstract}


\IEEEpeerreviewmaketitle
\section{Introduction}

In today's world, multimedia is an essential component in every sector. From education to industry, we need multimedia applications. It helps the user to understand abstract contents clearly and efficiently. We can not get a good understanding by looking at thousands of data. But if we look into only one graph representing these data, we can relate more with the dataset. Currently, we all use smartphones. People can capture unlimited images or videos using smartphones. For security purposes, CCTV cameras are installed everywhere by the users. We can process these data to get essential knowledge in a short time in a more efficient way. Multimedia techniques are a broad research field on their own. Here are some papers related to multimedia in different sectors that we reviewed and presented in this paper. We have categorized the various multimedia sector, and under each section, you will find the related research papers that we reviewed.

\section{Multimedia in Healthcare, Medicine and Education}

Multimedia has tremendous potential in the healthcare sector like  x-ray,magnetic resonance imaging (MRI), clinical audio-visual notes, online education etc. Researchers are working towards developing various multimedia algorithms and techniques to provide for better healthcare, education system and to complement the classic scenario with multimedia technology.
Some of the papers proposed by researchers in this domain are reviewed below.

\subsection{Airway Segmentation and Measurement in CT Images}
The various medical image registration techniques that then existed brought along with it several issues like low accuracy, limited applications and significant input processing. The multimodal techniques that were proposed later focused on lower airway trees but very few effective approaches existed for upper airway segmentation for CT and MR imaging.
In 2007 , A. Basu, I. Cheng et al. proposed a method to use Cone Beam CT image data to construct airways and measure their volume. This technique proved to be better than conventional CT in terms of scanning time, radiation dose and efficiency in the number of images taken. It also allowed for the evaluation of soft tissue characteristics.
The solution proposed involved Gradient Vector Flow (GVF) snakes and contour of an adjacent CT slice to track the current slice\cite{1}. Prewitt Edge Detection and snake-shifting using a heuristic approach was added to make the solution more robust. Prior knowledge of the shape of the airway can detect the airway in the first slice automatically. This is used for snake initialization of second slice and so on. The GVF snake was represented as shown below.

\begin{figure}[htbp]
\centerline{\includegraphics[width=6cm, height=1cm]{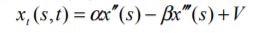}}
\caption{GVF snake representation. \cite{1}}
\end{figure}

The parametric curve solution for this equation was termed as GVF snake. The first term comprises of the first-order and second-order continuity term where the former has a large value where there is discontinuity in the curve and the latter has a large value where the curve has a sharp bend. $\alpha$ and $\beta$ control the stretching and bending of the contour at a point.An airway volume measurement was also proposed which aimed at keeping the distance between CT slices significantly small to produce accurate results.

\subsection{Automatic Segmentation of Spinal Cord MRI Using Symmetric Boundary Tracing}
In 2010, D. P. Mukherjee, I. Cheng, N. Ray, V. Mushahwar,M. Lebel, and A. Basu proposed a fully automatic technique to extract spinal cord from MRI and provide visualization. In this method, once the spinal cord is identified, image unskewing is performed by drawing a circle at the center of the mass of spinal cord region. The half circles that would be created by the axis of symmetry are then divided into N sectors and matched using the Bhattacharya coefficient (BC) between the intensity histograms of the sectors\cite{2}. The matching score was formulated as below. 
\begin{figure}[htbp]
\centerline{\includegraphics[width=8cm, height=2cm]{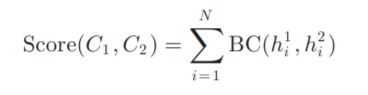}}
\caption{Matching score formula \cite{2}}
\end{figure}
Here  h$_i^1$ and h$_i^2$ indicate the intensity histograms of two corresponding sectors C1 and C2. The diameter that results in the highest score is considered as the symmetry axis. 

\begin{figure}[htbp]
\centerline{\includegraphics[width=7cm, height=1cm]{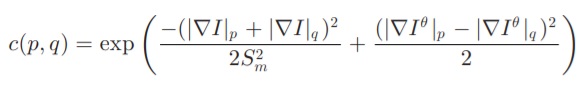}}
\caption{Cost Function for tracing a pixel\cite{2}}
\end{figure}
Further the muscle region is segmented from the unskewed image using active tracing algorithm as shown in Figure 3.
Here c(p,q) is the cost of tracing pixel (p,q) and |$\nabla$I| is the gradient magnitude and |$\nabla$I$^\theta$| is the gradient orientation for an image I. The approach then is to do a directed graph search for an optimal or shortest path in the energy surface defined by the cost function. The solution was enhanced to improve it's performance in the presence of image noise by evolving the active trace result along the normal direction of contour for the initial slice and use it to counter abnormalities in other slices. The slice which has the most symmetric active trace using shortest path energy minimization approach is used for initialization.

\subsection{Gradient vector flow based active shape model for lung field segmentation in chest radiographs}
In 2009, T. Xu, M. Mandal, R. Long, and A. Basu proposed a modified gradient vector flow-based active shape model (GVF-ASM) fitting algorithm for extracting lung field from chest radiographs. The authors proposed a new equation to reduce complexity and improve accuracy of searching as follows:
\begin{figure}[htbp]
\centerline{\includegraphics[width=6cm, height=1cm]{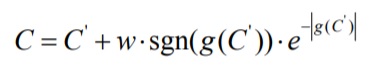}}
\caption{Points evolution equation. \cite{3}}
\end{figure}
Here the sgn function is for the GVF vector's direction and  the exponential function works as a smooth monotonically decreasing function. The shape model's parameters are updated within constraints once the control points are calculated and the stopping criteria would be the number of iteration times or a threshold of the point's Euclidean distance between two consecutive iterations. The proposed solution was compared with GVF-ASM segmentation for points evolution and it was observed that the former was more accurate and robust.

\begin{figure}[htbp]
\centerline{\includegraphics[width=8cm, height=5cm]{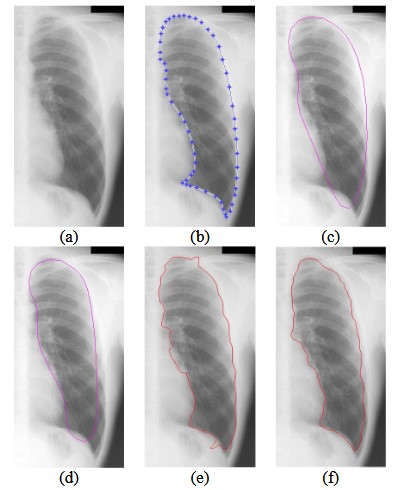}}
\caption{Performance of the proposed method for lung field segmentation (a)  Left  lung  image  for  test;  (b)  Manually  outlined  contour;  (c)  A PDM-based initialization using 20 training images; (d) ASM [14] result; (e) GVF-ASM [15] result; (f) Proposed method’s result. \cite{3}}
\end{figure}

\subsection{Interactive Multimedia for Adaptive Online Education}
In this paper \cite{17}, the authors discussed the possibilities of successfully delivering multimedia education by better concept representation than the current online education systems. Providing only multiple-choice questions or fill-in-the-blanks format by a system is not an efficient method for user engagement and user satisfaction. Keeping this in mind, the authors had designed CROME (Computer Reinforced Online Multimedia Education) framework to make the education more engaging by the users. They found that multimedia education was more effective because of its interactiveness and dynamic nature. Moreover, they discovered that abstract concepts could be easily representable using visual effects, animation, graphics, or 3D. Students love games, and they remained interested as multimedia can create a game-like environment. Visual effects help a student to understand a concept. Artificial intelligence (AI) can help to suggest a paper by understanding students' needs. AI can also understand a student's capability of understanding a particular concept. Using this information, AI can show more tutorials regarding the topic or set better exam questions. Their design framework is given below- 

\begin{figure}[htbp]
\centerline{\includegraphics[width=8cm, height=5cm]{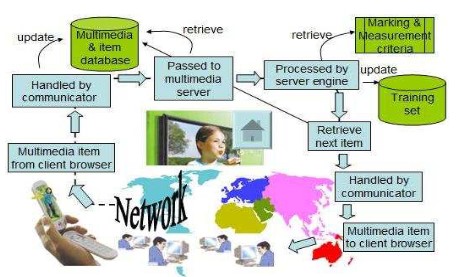}}
\caption{ The multimedia data flow in the CROME framework proposed by the authors \cite{17}}
\end{figure}

\section{ Human Facial Feature Detection, tracking, motion capture, Pose Recognition and Panoramic View}

Human facial features like eyes and nose, which are the most significant features contributing to different facial expressions, can be detected and tracked using techniques like template and texture mapping forr model based coding and synthesis of realistic facial expressions. Human pose recognition involves techniques like discrete Fourier transform, Radon transform (RT) and Hough transform(HT). Few papers related to this field of research are reviewed below:

\subsection{Eye tracking and Animation for MPEG-4 Coding}
A. Basu et al. proposed techniques for detection, tracking and modeling of eye movements along with animation models. This technique uses Hough Transform (HT) and deformable template matching along with color information to extract eyes\cite{4}. Iris is first detected using a gradient based HT for circles using a Sobel filter and performing non-maximum suppression. The position of a point on the upper and lower eyelids is detected by using horizontal integral projection. Eye feature tracking also involves similar steps with small differences. The eye features extracted from each frame of image sequence are used to synthesize the real motions of the eye on a 3D facial model. Real eye expression is then synthesized by using texture mapping technique on animated eye model and the original image.

\subsection{Nose Shape Estimation and Tracking for Model-Based Coding}
In 2001,A. Basu and L. Yin proposed a human facial feature detection method for low-bit rate model based coding. In this approach, a two-stage region growing method is used to localize facial feature regions on an extracted silhouette of a moving head\cite{13}. This helped in making feature detection less sensitive to noise.
Skin region growing was performed initially using global region growing and then the facial features are extracted using local region growing. Color information was used to solve the problem of initial localization in the deformable template matching technique that is used in this paper for nose shape detection and tracking. A geometric template looking like a pair curve with a leaf-like shape is applied on the nostril region and a pair of templates for right side and left side of the nose are applied for nostril and nose-side estimation\cite{13}. Finally, a 3D wireframe model is matched onto the individual face to track the motion of facial expression.

\subsection{Integrating active face tracking with model based coding}
This paper was published in 1999 by Linjun Yin and Anup Basu \cite{8}. In this paper, the authors discussed the problems regarding a talking face in front of an active camera. This active camera can be still or moving. But here, they have considered this active camera as moving and discussed possible difficulties for calculation. They have also proposed a solution that could track a talking face with an active camera in this paper. For their experiment, at first, they compensated the background of the successive frames using this paper's technique (Murray and Basu, 1994). Next, they used a morphological filter to reduce the noise. After that, they calculated consecutive frame differences to estimate the silhouette. This way, they detected the face region. Using deformable template technique and Hough transform, they extracted the facial features like nose, mouth, etc. Then they applied model fitting to animate the facial model. Their system only required input image frames, and the rest of the part was calculated and animated automatically by their model. To achieve detection, the motion of the speaker must need to be slow.

\subsection{
Perceptually Guided Fast Compression of 3D Motion Capture Data}
In this paper \cite{9}, the authors proposed a perception-based data compression technique for animations. In any animation video, when multiple dynamic characters are present, the bandwidth is typically high. Because the online software needs to handle the recording process as well as needs to translate it on a digital model. Therefore the authors came up with a solution, which could achieve a better compression ratio with a shorter compression and decompression time. They studied the different state-of-the-art approaches and discovered that the Wavelet coding was the best because of its processing and incorporating human perceptual factors. To locate the high attention region, they applied Interactivity-Stimulus-Attention Model (ISAM). They focused on bone lengths and variation in the rotation of bones. Based on their understanding from different researches, they build an algorithm on the lossy compression technique. Thus they achieved their desirable compression ratio.

\subsection{Panoramic Video with Predictive Windows for Telepresence Applications }
In this paper \cite{10}, the authors described an application of the predictive Kalman filter to display panoramic images. Panoramic images can be beneficial in terms of remote works (construction, robotics, etc.). Human operators highly rely on their visual perception for decisions. Thus the bandwidth needs to be sufficient on both the local and remote sites. Also, for the video or image frame transmission, the camera needs to cover a wide range of areas, which is not possible using one single camera. So they have proposed a method of taking a panoramic image using a CCD camera with a conic mirror to cover a 360-degree angle of the area. For better visualization for the operator, the authors proposed a rectangular window with increased size to cover the 360-degree area. Finally, they used the Kalman filter to predict the viewing direction while displaying panoramic images.

\subsection{Pose Recognition using the Radon Transform}
In this paper \cite{11}, the authors proposed a method for recognizing human pose using radon transform. Their methodology depends on two assumptions -

\begin{itemize}
  \item Any pose of arms or legs should be replaced by a medial skeleton representing the pose.
  \item Representation of that pose in Radon Transform space should provide classification information of that pose.
\end{itemize}

At first, they separated the foreground from the background by using the statistical background modeling approach. Next, to get the medial skeleton, a thinning procedure was applied. Radon Transform is then applied to get the skeletal orientation lines. After that, the authors modified the Spatial Maxima Mapping Algorithm to work fast and applied. This algorithm generated some numerical values. Based on the values, the pose was estimated by the authors. Their proposed approach is computationally expensive.

\subsection{A Multisensor Technique for Gesture RecognitionThrough Intelligent Skeletal Pose Analysis}
Because of the latest sensor technology and computer vision techniques, it is possible to recognize hand gestures. However, due to occlusion of fingers caused by quick movements or other parts of the hands makes it difficult to determine the hand gesture accurately in real-time. To solve this problem, the authors of this paper \cite{16} came up with a solution. They used multisensor technique to determine the hand gestures in real-time and increased the pose estimation accuracy. They also used pose estimation from multiple sensors to overcome the occlusion issue. These sensors were set up at different angles so that they can detect the fingers. Each of the sensors is capable of determining the hand position skeleton. They build an offline model which could select an appropriate subset of the skeleton pose estimation parameters independently in real-time. During the experiment, dual physical sensors were placed in the room by the authors. Then they analyzed the accuracy when the hands were placed near the sensors by the examiner at different angles, with controlling poses of the fingers (for example, tap, pinch, and open hand position). In this way, they calculated the accuracy for a single sensor and dual sensors. Finally, they placed artificial hands to understand the accuracy and errors more efficiently.

\section{ Multimedia in Data Handling and Processing}

\subsection{Variable Resolution Teleconferencing}
A. Basu et al. proposed a prototype for a teleconferencing system based on Variable Resolution (VR) sensing\cite{6}. This approach used multiple scaling factors to make the task of mapping transformed images to a rectangular storage field easy in order to achieve full space utilization. The use of multiple foveas proved to be beneficial in situations where there is more than one area of interest to the observer. Straight pixel sampling was used for interpolation where one pixel is used to represent it's nearest neighbours. VR was compared to JPEG w.r.t the compression time, compression ratio and resulting image quality. It was observed that VR transformation was extremely fast but did not perform better than JPEG when it came to image quality and compression ratio. Modifications to the proposed VR algorithm were then made to enhance it's performance by changing the ways compressed image is generated, decompressed and the nearest neighbor method used. The prototype proposed was ideal for teleconferencing market especially for machines with limited speed with an added advantage of the frame rate being maintained without additional hardware.
\begin{figure}[htbp]
\centerline{\includegraphics[width=8cm, height=5cm]{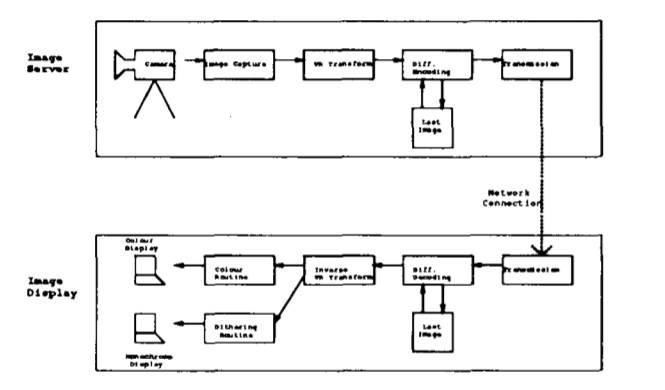}}
\caption{Videoconferencing system prototype. \cite{6}}
\end{figure}

\subsection{Stereo Matching Using Random Walks}
In 2008, R. Shen, I. Cheng, X. Li, and A. Basu proposed a novel two-phase stereo matching algorithm for color images using random walks framework\cite{5}. The parameters used were adapted automatically according to the illumination condition of the input image pair. This approach takes stereo pair of input images and computes the initial disparity maps. A set of reliable matching pixels are obtained using left - right checking. A final disparity map is computed after performing texture-less region handling. The matching is based on the prior matrices and Laplacian matrices on the information of adjacent pixels. Then the model
uses  to determine the disparities of unreliable regions are determined using the the set of pixels as seeds and solving a Dirichlet problem. The lightning
and illumination changes are also taken into consideration to achieve higher accuracy while building the disparity maps. Noise reduction is also performed simultaneously in between all these steps.

\subsection{Videoconferencing using Spatially Varying Sensing with Multiple and Moving Foveae}
In the paper \cite{7}, A. Basu et al. proposed a videoconferencing method using the concept of spatially varying sensing along with a fast videoconferencing prototype for desktop computers, The concept of variable resolution(VR) combined with varying spatial resolution has also been proposed for scenarios having more than one area of interest in images. Under VR transform, a pixel with polar coordinates (r, theta)..
VR image compression leads to one pixel representing several pixels in original image to which Bilinear interpolation is applied mostly for decompression. Sometimes the images are not rectangular but the storage field is. Two methods have been proposed to solve this problem. The first solution uses multiple scaling factors to transform image for full space utilization. The second approach is comparatively less complex and involves isolating the vertical and horizontal components. In the case of moving foveas , a Look Up Table (LUT) is used whose entries are the results of the transformation for all points lying in the first quadrant, when the fovea is at the bottom left corner of the image. Information pertaining to first quadrant in LUT can be used to obtain VR transform for other quadrants. Cooperative and Competitive foveae approaches have been analyzed for cases having multiple foveae.

\subsection{Variable-resolution character thinning}
This paper was published in 1991 by Xiaobo Li and Anup Basu. For character recognition, character thinning is a pre-processing step. People used character thinning to get the skeleton of the characters. At that time, the character thinning algorithm used only small windows, which lead to shape distortion and misclassification of the character. To solve that issue, the authors proposed an algorithm. That algorithm used a large 9*9 window. At the center of that large window, there was a 3*3 small window. The smaller window was using a higher resolution, where its peripheral area inside the bigger window was lower in resolution. The human visual system also works the same. It has a high-resolution fovea and a low-resolution peripheral vision. They took this approach to thin the character pixels. Furthermore, they used Guo's algorithm A1 for their experiment as they found this algorithm was better in terms of accuracy at that time. Finally, they achieved their desired goals.

\begin{figure}[htbp]
\centerline{\includegraphics[width=9cm, height=4cm]{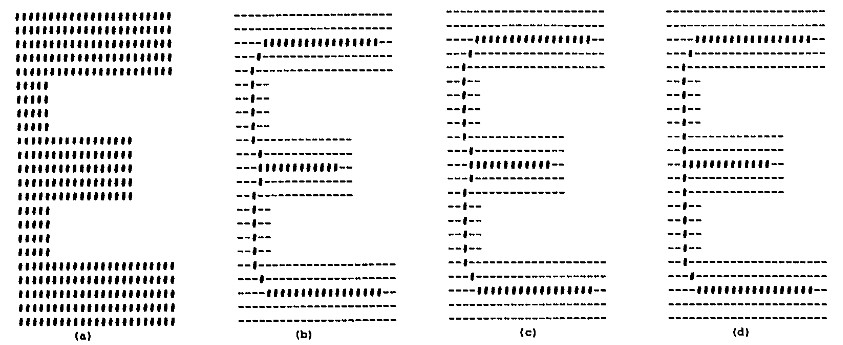}}
\caption{Input character and thinning results of three algorithms: (a) input, (b) Holt's method, (c) Guo's method, (d) proposed method. \cite{12}}
\end{figure}

\section{Conclusion}
This survey aims at summarizing some valuable research done on various multimedia technologies and algorithms in areas like healthcare, human facial feature detection and tracking, multimedia data processing. We have highlighted the key-points of the solutions proposed by the researchers along with the enhancements made to them to produce more efficient and robust algorithms. We hope to continue exploring such methods and apply the knowledge of computer vision to multimedia data to solve daily life problems efficiently.

\end{document}